\documentclass[12pt]{article}
\usepackage{mathrsfs}
\begin{document} 
\title{\bf D3/D7 Inflation in a Type-0B String Background}
\author{{\sc Huan-Xiong Yang \thanks{E-mail: hxyang@zimp.zju.edu.cn}}
\\
{~~}
\\
{\it Zhejiang Institute of Modern Physics, Physics Department,}
\\
{\it Zhejiang University, Hangzhou, 310027, P. R. China }
\\
{~~}
\\
{~~}}

\date{\today}
\maketitle

\begin{abstract}
The D3/D7 inflation is studied in the framework of Type 0B string theory. Due to the presence
of the closed string tachyon, the constant gauge field flux on the transverse worldvolume of D7-brane
is almost arbitrary in having an attractive inflaton potential energy. Besides, a positive cosmological
constant appears in this model whose magnitude is related to the volume modulus of the D7-brane
transverse worldvolume.
\\
{~~}
\\
{~~}
\\
{\bf keywords:}~~~{Type-0B string, D-branes, Mirage Inflation}
\end{abstract}

\maketitle

\newpage

The progress of superstring theory has brought out an crazy idea that our observable universe is
effectively a 4-dimensional brane embedded in a higher dimensional spacetime. The idea not only
provides probable answer to the long standing hierarchy problem in the standard Model of particle
physics\cite{RS, RS2}, it does also open a way for understanding the cosmology of our universe, in particular
the inflationary phase in its early evolution. Brane inflation in string theory
framework has intensively been investigated these years in various contexts such as\cite{KKLT, KKLMMT,
Kal, Das, Burg, Brodie, Hsu, HK, Cline, Pappa1, Pappa2}, among which the D3/D7 model proposed by
Dasgupta, Herdeiro, Hirano and Kallosh in Ref.\cite{Das} appears very attractive. In the scenario
of Type IIB string theory, the D3/D7 inflation model has an effective description as a hybrid
inflation or as a D-term inflation model\cite{Bin, Dvali, Kal2}. The flat direction of the inflaton
potential is associated with the shift symmetry which protects the flatness of inflaton potential
and then the expected small inflaton mass even with the stabilization of some string moduli\cite{HK}.
Recently, a generalized version of the D3/D7 model has been developed to eliminate the production
of the topologically stable Abrikosov-Nielsen-Olesen (ANO) cosmic strings\cite{Das2, Davis}.
This might be an important achievement because the ANO cosmic strings produced at the end of
original D3/D7 inflation lead to additional CMB anisotropy that is in general inconsistent with
the results of WMAP observation.

The D3/D7 inflation model consists of a mobile D7-brane moving in the background of a stack of heavy D3-branes.
The distance between them which is an open string modulus is interpreted
as the inflaton field. In Type IIB string theory scenario the supersymmetry breaking parameter is
related to the presence of a constant antisymmetric gauge field flux $F_2$ on the sub-worldvolume of
the D7-brane which is transverse to the D3-branes. When the flux is not self-dual in this 4-dimensional
compact space, the supersymmetry of the combined brane system is broken. In the present paper, we suggest
an alternative D3/D7 inflation model in framework of Type-0B string theory\cite{Kleb}. The novel
feature of Type-0B D3/D7 model is the mechanism for having an attractive inflaton potential energy.
One remarkable feature of Type-0 string theory is that there exists a tachyon field in its closed
string sector, which signs the breakdown of supersymmetry. Due to this closed string tachyon, in our
D3/D7 model a slow-roll inflationary stage would take place even if the constant gauge flux $F_2$ on
the D7-brane is self-dual. Besides, our construction provides a geometric explanation for the positiveness and
smallness of the 4-dimensional cosmological constant which depends on the volume integration of
the D7-brane action over the transverse space. This implies that the interaction between D3/D7 branes lifts
the minimum of inflaton potential energy to a 4D de Sitter space.

The Type-0B string theory has no spacetime fermions in its spectra. The massless bosonic fields are as in the Type IIB theory
but with the doubled sets of the RR fields. In string frame, the low-energy effective action of Type-0B supergravity plus
D3/D7 branes reads\cite{Kleb},
\begin{equation}\label{eq: 1}
\left. \begin{array}{lll} S_{10} & = & \int d^{10}x \sqrt{-g} ~\bigg[~e^{-2\phi} (R +4\partial^{\mu}\phi
\partial_{\mu}\phi -\frac{1}{4}\partial^{\mu}T\partial_{\mu}T -\frac{1}{4}m^2T^2 -\frac{1}{12}H_{3}^2)\\
&   & -\frac{1}{2}\Big(1+\frac{T^2}{2}\Big)(|F_3|^2+|\bar{F}_3|^2)~-|F_3 \bar{F}_3|T  -\frac{1}{2}\Big(1+T+\frac{T^2}{2}\Big)|F_5|^2~ \\
&   & +{\mathscr L}_{D3}+{\mathscr L}_{D7}\bigg]
\end{array}
\right.
\end{equation}
where $\phi$ and $T$ are respectively the dilaton field and closed string tachyon in NS-NS sector, $H_3$ is the
strength of NS-NS B-field. $F_3$, $\bar{F}_3$ and $F_5$ are the fluxes associated with the RR potentials.
Different from the RR flux $F_5$ in Type IIB string theory, the Type-0B field strength $F_5$ is not necessary
to be self-dual\cite{Kleb}. There is in addition terms in Eq.(\ref{eq: 1}) describing the coupling between the
RR fields and the closed string tachyon, among which the terms such as $|F_5|^2T^2$ provide contributions to
shift the effective $(\textrm{mass})^2$ of the tachyon field $T$ to positive, giving an appealing mechanism
for stabilizing the background. ${\mathscr L}_{D3}$ and ${\mathscr L}_{D7}$ are Lagrangians of D3 and D7 branes
respectively. The 3-form fluxes $F_3$, $\bar{F}_3$ and $H_3$ are necessary for fixing all of the string moduli (except the volume modulus) when we consider the compactification of this type-0B string background\cite{GKP, Burg}. We suppose that
such moduli have been fixed and choose to set $H_3=F_3=\bar{F}_3=0$ in the following. Besides, we adopt probe approximation\cite{Kiritsis} so that ${\mathscr L}_{D7}$ can be ignored in determining the bulk fields. The D3 branes are assumed to be so heavy that ${\mathscr L}_{D3}$ can also be ignored in Eq.(\ref{eq: 1})\cite{Brodie}, which demonstrate
their presence by the RR potential $C_4$ that couples to the D3 brane worldvolume. The equations of motion of the bulk
fields in this background are as follows,
\begin{equation}\label{eq: 2}
\left.
\begin{array}{l}
R_{\mu \nu}+4\partial_{\mu}\phi \partial_{\nu}\phi -\frac{1}{4}\partial_{\mu}T \partial_{\nu}T -\frac{m^2}{32}
g_{\mu \nu}T^2 -\frac{f(T)}{4\cdot 4!}e^{2\phi}\big( F_{\mu \xi_2 \cdots \xi_5}F_{\nu}^{~\xi_2 \cdots
\xi_5}-\frac{1}{10}g_{\mu \nu}F^2_5 \big) = 0~,\\
\frac{1}{\sqrt{-g}}\partial_{\mu}(\sqrt{-g}\partial^{\mu}\phi)-2\partial_{\mu}\phi \partial^{\mu}\phi
-\frac{1}{16}m^2T^2=0~,\\
\frac{1}{\sqrt{-g}}\partial_{\mu}(\sqrt{-g}\partial^{\mu}T)-2\partial_{\mu}\phi \partial^{\mu}T -m^2T
-\frac{1}{2\cdot 5!}e^{2\phi}f^{\prime}(T)F^2_5=0~,\\
\partial_{\mu}\big[\sqrt{-g}f(T)F^{\mu \xi_2 \cdots \xi_5} \big]=0~,
\end{array}
\right.
\end{equation}
where,
\begin{equation}\label{eq: 3}
f(T)=1+T+\frac{1}{2}T^2~.
\end{equation}

The appearance of tachyon field $T$ in Eqs.(\ref{eq: 2})makes these equations too complicated to
be solved exactly\cite{Pappa1, Pappa2, Kleb}. We suppose that in the equation of motion of the
tachyon field the term $\frac{1}{2\cdot 5!}e^{2\phi}F_5^2$ dominates the potential part. In this
case, we can ignore in Eqs.(\ref{eq: 2}) the terms proportional to $m^2$. The tachyon field $T$
can be regarded as a constant background near the condensation point [at which $f^{\prime}(T)=0$].
Under such an approximation and the assumption that the volume modulus has been stabilized by the
nonperturbative D3/D7 brane interaction\cite{Burg}, Eqs.(\ref{eq: 2}) have the following static,
spherically symmetric solution
\begin{equation}\label{eq: 4}
\left.
\begin{array}{l}
 ds^2=H^{-\frac{1}{2}}(-dt^2 +\sum_{i=1}^{3}dx_i^2) +H^{\frac{1}{2}}(dr^2 + r^2d\Omega_5^2)~,\\
 \phi=0~,\\
 C_4=\frac{Q}{f(T)L^4}(H^{-1}-1)dt\wedge dx^1\wedge dx^2\wedge dx^3~,\\
 F_5=\frac{4Q}{f(T)H^2r^5}dt\wedge dx^1\wedge dx^2\wedge dx^3\wedge dr~,
\end{array}
\right.
\end{equation}
where the warp factor reads,
\begin{equation}\label{eq: 5}
H=1+({L}/{r})^4~.
\end{equation}
The constants $L$ is related to the RR charge $Q$ of the D3-brane by equation $L^4=\frac{Q}{\sqrt{2f(T)}}$.

We now study the  dynamics of our universe D7-brane in the above Type-0B plus D3-brane background.
The D7-brane is regarded as a probe when it moves along the geodesic. Relying on the appearance of
closed string tachyon $T$, it is plausible to expect the attractive force between these two
kinds of branes. As the D7-brane moves, the induced world-volume metric becomes a function of time,
so it might inflate during its motion through a bulk spacetime. In this way, D3/D7 model shares some
of the features of the mirage cosmology\cite{Kiritsis, Kiritsis2}. The D3/D7-brane system is supposed
to have a spatial configuration in Table 1 where a constant gauge field flux $F_2$ is added along
the transverse directions on the worldvolume of D7-brane. In ambient space, the heavy D3-brane is
at $(x^4)^2+(x^5)^2=0$ while D7-brane is initially at some ${\psi}^2=(x^4)^2+(x^5)^2 \gg {\psi}^2_c$,
where ${\psi}^2_c$ stands for some critical distance in between beyond which an open string tachyon
would develop.  The dynamics of the probe brane is governed by the following Dirac-Born-Infled action
\begin{table}\label{eq: 8}
\begin{center}
\begin{tabular}{|c|c|c|c|c|c|c|c|c|c|c|}
\hline \multicolumn{1}{|c|}{\textsf{~~~}} & \multicolumn{1}{|c|}{{\textsf{0}~}}
&\multicolumn{1}{|c|}{{\textsf{1}~}} & \multicolumn{1}{|c|}{{\textsf{2}~}} &\multicolumn{1}{|c|}{{\textsf{3}~}}
&\multicolumn{1}{|c|}{{\textsf{~4}~}} &\multicolumn{1}{|c|}{{\textsf{~5}~}} &\multicolumn{1}{|c|}{{\textsf{6}~}}
&\multicolumn{1}{|c|}{{\textsf{7}~}} &\multicolumn{1}{|c|}{{\textsf{8}~}}
&\multicolumn{1}{|c|}{{\textsf{9}~}}\\
\hline D3 & $\times$ & $\times$ & $\times$ & $\times$ & & & & & & \\
\hline D7 & $\times$ & $\times$ & $\times$ & $\times$ & & & $\times$ & $\times$ & $\times$ & $\times$  \\
\hline $F_2$ &  &  &  &  &  &  & $\times$ & $\times$ & $\times$ &  $\times$    \\
\hline
\end{tabular}
\end{center}
\caption{\textsf{Brane Configuration in Type-0 String Theory}}

\end{table}
\begin{equation}\label{eq: 9}
S_7=-T_7\int d^8\xi e^{-\phi}\sqrt{-\det(G_{\alpha \beta} +2\pi \alpha^{\prime}F_{\alpha \beta})} + T_7 \int
\sum C_{p+1}\wedge e^{F_2}
\end{equation}
where $T_7(>0)$ is the tension of D7 brane and $G_{\alpha \beta}$ the induced metric on its worldvolume
\begin{equation}\label{eq: 10}
G_{\alpha \beta}=g_{\mu \nu}\frac{\partial x^{\mu}}{\partial \xi^{\alpha}}\frac{\partial x^{\nu}}{\partial
\xi^{\beta}}, ~
\end{equation}
The constant gauge field flux $F_2$ on the worldvolume of D7-brane takes the form
\begin{equation}\label{eq: 11}
F_2=\frac{1}{2\pi \alpha^{\prime}}(\tan\theta_1 d\xi^6\wedge d\xi^7 +\tan\theta_2 d\xi^8\wedge
d\xi^9)
\end{equation}

We fix the gauge freedom in action (\ref{eq: 9}) by choosing the so-called static gauge, $\xi^{\alpha}=x^{\alpha}
~(\alpha=0,~1,~2,~3,~6,~7,~8,~9)$. Let $\frac{L^2}{r^2}=\sigma^2+\psi^2$.  In the regime $r^2\ll L^2$
(corresponding to large values of field $\psi$) the Type-0B supergravity metric (\ref{eq: 4}) that is
approximately of $AdS_5\times S^5$ induces the following metric on the D7-brane worldvolume
\begin{equation}
    \label{eq: 12a}
    \left.
    \begin{array}{lll}
        G_{00}& = & -(\sigma^2+\psi^2)^{-1}\big[1-(\sigma^2+\psi^2)^{-1}{L^2\psi^2{\dot{\psi}}^2}\big] \\
        G_{ij}& = & \delta_{ij}(\sigma^2+\psi^2)^{-1}, ~\,\,\,\,\,\,\,(i,j=1,\, 2, \, 3)\\
        G_{ab}& = & \delta_{ab}(\sigma^2+\psi^2),~\,\,\,\,\,\,\,\,\,\,\, (a, b=6, \,7, \,8,\,9)
    \end{array}
    \right.
\end{equation}
Substituting metric (\ref{eq: 12a}) into Eq.(\ref{eq: 9}) and finishing the integration over the compact
sub-worldvolume of the D7-brane, we can get an effective 4-dimensional action for the field $\psi$ in
our universe. We denote the volume modulus of the D7-brane transverse sub-worldvolume by a ``cut-off'' $\Lambda$.
The 4-dimensional effective action of the scalar field $\psi$ is then $S_{\psi}=\int d^4x {\mathscr L}_{\emph{eff}}$~,
with
\begin{equation}
    \label{eq: 13a}
    {\mathscr L}_{\emph{eff}}=\frac{1}{2}(T_7\pi^2 L^2)\psi^2[\Lambda-\psi^2
    \ln(1+\frac{\Lambda}{\psi^2})]{\dot{\psi}}^2 -V(\psi)~.
\end{equation}
The potential energy of $\psi$ reads,
\begin{equation}
    \label{eq: 14a}
    V(\psi)=\lambda -\frac{1}{2}(\pi^2T_7 \zeta) \Big[\frac{\Lambda}{\Lambda+\psi^2}- \ln (1+
    \frac{\Lambda}{\psi^2})\Big]~,
\end{equation}
where $\lambda= \frac{1}{2}(\pi^2 T_7 \Lambda^2)(1+\sqrt{\frac{2}{f(T)}} \tan\theta_1 \tan \theta_2)$ and
\begin{equation}
        \label{eq: 15a}
        \zeta = (\tan \theta_1 - \tan \theta_2)^2 + 2 \Bigg[~1-\sqrt{\frac{2}{f(T)}}~ \Bigg]
        \tan \theta_1 \tan \theta_2 ~.
\end{equation}
The constant $\lambda$ plays a role of the 4-dimensional cosmological constant while the parameter $\zeta$
determines whether the interaction between D7/D3 branes is of attractive. Take the limiting case $\Lambda \gg \psi^2$
as an example. The potential (\ref{eq: 14a}) has approximately a logarithmic dependence on the ``inflaton'' $\psi$
\begin{displaymath}
    V(\psi) \sim -\frac{1}{2}(\pi^2T_7 \zeta)\ln \big({\psi^2}/{\Lambda} \big)~.
\end{displaymath}
When $\zeta <0$, it is similar to the attractive potential of the inflaton in the hybrid inflation model
or in the D-term inflation models of the supergravity theory\cite{Bin, Dvali, Kal2, Davis}. Being negative
for $\zeta$ in the considered Type-0B scenario is apparently possible if the inequality $0<f(T)<2$ holds.
If the closed string tachyon is at its condensation point, $f(T)=1/2$, the attractiveness of the potential
imposes a constraint $$(\tan \theta_1 - \tan \theta_2)^2<2\tan \theta_1 \tan \theta_2$$ on the constant
gauge field flux $F_2$. Obviously, that $F_2$ is self-dual over the transverse space ($\theta_1=
\theta_2~ {\not =} 0$) does not spoil this requirement.

In this paper we examine the possibility of mirage inflation on the D7-brane worldvolume in the contrary
limiting case $\Lambda \ll \psi^2$. At first, however, let us derive the Friedmann equation and the
equation of motion of the inflaton $\psi$ in a general situation. The full action on the
uncompactified sub-worldvolume of the D7-brane is written as\cite{KKLMMT, Brodie}:
\begin{equation}
        \label{eq: 16a}
        S=\int d^4x \sqrt{-g}\Big(\frac{M_{\textup {p}}^2}{2}R + {\mathscr L}_{\emph{inflaton}}\Big)~.
\end{equation}
The expression of ${\mathscr L}_{\emph{inflaton}}$ is almost the same as that of ${\mathscr L}_{\emph{eff}}$
in Eq.(\ref{eq: 13a}) except that ${\dot{\psi}}^2$ should be replaced by $-g^{\mu \nu}\partial_{\mu}\psi
\partial_{\nu}\psi$. The 4-dimensional metric fields satisfy the standard Einstein equations $R_{\mu \nu}=
\frac{1}{M^2_{\textup p}}(T_{\mu \nu}-\frac{1}{2}g_{\mu \nu}T)$, with the following contribution of the scalar field $\psi$
to the energy-momentum tensor:
\begin{equation}
        \label{eq: 17a}
        \left.
        \begin{array}{l}
            T_{\mu \nu}  =  T_7 (\pi L)^2 \psi^2 \big[\Lambda - \psi^2 \ln \big(1+
            \frac{\Lambda}{\psi^2}\big)\big]\partial_{\mu}\psi \partial_{\nu}\psi
            + g_{\mu \nu}{\mathscr L}_{\emph{inflaton}}\,\,.\\
        \end{array}
        \right.
\end{equation}
The equation of motion of inflaton field $\psi$ reads,
\begin{equation}
        \label{eq: 18a}
        \left.
        \begin{array}{ll}
        0 = & \psi^2\big[ \Lambda - \psi^2 \ln \big(1+ \frac{\Lambda}{\psi^2}\big)\big]
        \frac{1}{\sqrt{-g}}\partial_{\mu}(\sqrt{-g}g^{\mu \nu}\partial_{\nu}\psi)
        +\psi \big[\Lambda - 2\psi^2 \ln \big(1+ \frac{\Lambda}{\psi^2} \big) \big](\partial \psi)^2\\
        & +  \frac{\zeta \Lambda^2}{L^2 \psi (\Lambda + \psi^2)^2}~.
        \end{array}
        \right.
\end{equation}

It is common to take the inflaton $\psi$ to be spatially homogeneous and to assume the perfect fluid form for
its energy-momentum tensor\cite{Linde}. Under the assumption that the effective metric on the universe brane is of the
Friedmann-Robertson-Walker form
\begin{equation}
        \label{eq: 21a}
        ds^2_4=-(dt)^2 + a^2(t)\sum^{3}_{i=1}(dx^i)^2
\end{equation}
we can identify from Eq.(\ref{eq: 17a}) the energy density and pressure due to inflaton $\psi$:
\begin{equation}
        \label{eq: 22a}
        \left.
        \begin{array}{l}
            \rho  =  \frac{T_7(\pi L)^2 }{2} \psi^2 \big [ \Lambda - \psi^2 \ln \big(1+
            \frac{\Lambda}{\psi^2}\big)\big] {\dot{\psi}}^2 + V(\psi) \\
            P  =  \frac{T_7(\pi L)^2 }{2}\psi^2 \big [ \Lambda - \psi^2 \ln \big(1+
            \frac{\Lambda}{\psi^2}\big)\big] {\dot{\psi}}^2 - V(\psi) \\

        \end{array}
        \right.
\end{equation}
where the potential energy $V(\psi)$ is given in Eq.(\ref{eq: 14a}). As expected, Einstein equations
become the Friedmann equation ${\cal H}^2={\rho}/{3M^2_{\textup p}}$ with the
Hubble parameter defined as ${\cal H}={\dot{a}}/{a}$. Besides, the state equation of inflaton
can be written as $P=\omega \rho$. Inflation would be probable if $\omega <-\frac{1}{3}$, which
implies $T_7(\pi L)^2 \psi^2 \big [ \Lambda - \psi^2 \ln \big(1+\frac{\Lambda}{\psi^2}\big)\big]
{\dot{\psi}}^2 < V(\psi)$. The definite positiveness of parameter $\lambda$ enables us to suppose
that in Eq.(\ref{eq: 14a}) the cosmological constant term dominates the potential energy. Consequently,
the condition for inflation is recast as
\begin{equation}
        \label{eq: 22b}
        \psi^2 \big [ \Lambda - \psi^2 \ln \big(1+\frac{\Lambda}{\psi^2}\big)\big] {\dot{\psi}}^2
        < \frac{\Lambda^2}{2L^2} \Bigg[ 1+ \sqrt{\frac{2}{f(T)}} \tan \theta_1 \tan \theta_2 \Bigg]~.
\end{equation}

The positiveness of $\lambda$ combined with the fact that the inflaton potential is of the attractive implies that
our 4-dimensional universe on the D7-brane has a metastable de Sitter-like vacuum. The smallness of
cosmological constant from the cosmology observation data and its expression in the considered scenario suggest
that the inflationary phase of our universe model should correspond to the case $\Lambda \ll \psi^2$. It appears that
the contrary case $\Lambda \gg \psi^2$ corresponds to a vanishing cosmological constant and one has to introduce
an additional anti-D3-brane to lift the vacuum energy\cite{Burg}. We define
a new inflaton field $\varphi = \sqrt{\frac{T_7}{2}}(\pi L \Lambda)\psi$.  The Friedmann equation and the equation
of motion of this new inflaton in the limiting case $\Lambda \ll \psi^2$ turn out to be
\begin{equation}
        \label{eq: 23a}
        {\cal H}^2= \frac{V(\varphi)}{3M^2_{\textup p}},\,\,\,\,\,\,\,\,\,
        3{\cal H}\dot{\varphi}+V^{\prime}(\varphi)=0
\end{equation}
where,
\begin{equation}
        \label{eq: 24a}
        V(\varphi)= \lambda - \frac{\alpha}{\varphi^4}
\end{equation}
with $\alpha=-\frac{1}{16}T^3_7(\pi \Lambda)^6L^4\zeta$.  As mentioned previously, we have to impose the
constraint $\zeta<0$ to ensure $V(\varphi)$ being an attractive potential.

From the potential (\ref{eq: 24a}) we can calculate the slow-roll parameters
\begin{equation}
        \label{eq: 25a}
        \left.
        \begin{array}{l}
            \epsilon = \frac{1}{2}M^2_{\textup p}(V^{\prime}/V)^2 \approx 8M^2_{\textup p}
            \frac{\alpha^2}{\lambda^2 \varphi^{10}}~, \\
            \eta = M^2_{\textup p} (V^{\prime \prime}/V) \approx -20M^2_{\textup p}
            \frac{\alpha}{\lambda \varphi^6}~.
        \end{array}
        \right.
\end{equation}
In order for the slow-roll approximation to be valid, the inflaton must be on a regime of potential which
satisfies $\epsilon \ll 1$ and $|\eta| \ll 1$, of which the condition $|\eta| \ll 1$ is more restrictive in
string cosmology. In the considered case, $|\eta| \ll 1$ can be met by taking
\begin{equation}
        \label{eq: 26}
        \varphi \gg \bigg (20 M^2_{\textup p}
        {\alpha}/{\lambda} \bigg)^{\frac{1}{6}}.
\end{equation}
The amount of inflation is measured by the number of e-foldings
\begin{equation}
        \label{eq: 27}
        N=\ln \Bigg[ \frac{a(t_{\textup {end}})}{a(t)}\Bigg] = \int^{t_{\textup {end}}}_{t} {\cal H}dt
        =\frac{1}{M^2_{\textup p}} \int \frac{V}{V^{\prime}} d\varphi \approx
        \frac{\lambda \varphi^6}{24 M^2_{\textup p} \alpha}~.
\end{equation}
$\eta$ can be expressed as $\eta =-\frac{5}{6N}$ in terms of the e-folding number $N$.  Setting $N\approx 60$
gives $\eta \approx -0.014$.

The magnitude of the cosmological constant $\lambda$ in the present model can be estimated by the knowledge of the
slow-roll parameter $\epsilon$ and the adiabatic density perturbation $\delta_H$. The latter is defined as
\begin{equation}\label{eq: 28}
    \delta_H=\frac{1}{5\sqrt{3}\pi M^3_{\textup p}} \cdot \frac{V^{3/2}}{V^{\prime}} \approx \frac{1}{5\pi \sqrt{6\epsilon}}\bigg(\frac{\lambda}{M^4_{\textup p}}\bigg)^{\frac{1}{2}}~.
\end{equation}
If the perturbation $\delta_H$ is responsible for the structure of the observational universe, it should be
$1.9 \times 10^{-5}$ at $N \approx 60$. We take $|\eta| \gg \epsilon$ and use the value of $\epsilon$ given in
Ref.\cite{KKLMMT}, $\epsilon \approx 1.54 \times 10^{-11}$. Therefore,
\begin{equation}
        \label{eq: 29}
        \frac{\lambda}{M^4_{\textup p}} \approx 8.23 \times 10^{-18}~.
\end{equation}
Moreover, the tilt parameter is given by
\begin{equation}
        \label{eq: 30}
        n=1-6\epsilon+2\eta \approx 0.97
\end{equation}
which is in excellent agreement with the observational data from WMAP.

In conclusion, we have shown that in the Type-0B D3/D7 brane system our universe D7-brane could have an
slow-roll inflationary stage in its early cosmological evolution both in the limiting cases of $\Lambda \ll \psi^2$
and of $\Lambda \gg \psi^2$. The nonvanishing constant gauge field flux $F_2$ on the transverse subvolume of
D7-brane is necessary for having an inflaton potential in the D3/D7 inflationary models. In Type-0B scenario,
however, it is not constrained to violate the self-dual symmetry. It is the coupling of nonvanishing flux $F_2$
and the closed string tachyon that induces the probable inflationary evolution on the 4D extended
sub-worldvolume of D7-brane universe. Furthermore, the positive signature and the small magnitude of the
observed cosmological constant can have a stringy interpretation in this Type-0B D3/D7 inflationary model
in the case $\Lambda \ll \psi^2$.

\newpage
\section*{Acknowledgments}
The author thanks M.X. Luo, Q.P Su  and B. G. Cai for discussions. This work is partially supported by
CNSF-10375052 and the Startup Foundation of the Zhejiang Education Bureau.

\end{document}